\documentclass[useAMS,usenatbib]{mnras}
\usepackage{pifont}
\usepackage{graphicx}
\usepackage{amssymb}
\usepackage{amsmath}
\usepackage{abbrevs}
\usepackage{xspace}
\usepackage{float}
\usepackage[usenames,dvipsnames]{xcolor}
\usepackage{newtxtext,newtxmath}
\begingroup
  \catcode`\_=\active
  \gdef_#1{\ensuremath{\sb{\mathrm{#1}}}}
\endgroup
\mathcode`\_=\string"8000
\catcode`\_=12

 \newcommand{\rev}{\textcolor{Black}}

\newcommand{\Msolarpc}{M$_{\odot}$ / pc$^{2}$\xspace}

\newcommand{\atcc}{cm$^{-3}$\xspace}

\def\setsymbol#1#2{\expandafter\def\csname #1\endcsname{#2}}
\def\getsymbol#1{\csname #1\endcsname}

\def\Planck{\textit{Planck}}

\def\all2013resultspapers{\nocite{planck2013-p01, planck2013-p02, planck2013-p02a, planck2013-p02d, planck2013-p02b, planck2013-p03, planck2013-p03c, planck2013-p03f, planck2013-p03d, planck2013-p03e, planck2013-p01a, planck2013-p06, planck2013-p03a, planck2013-pip88, planck2013-p08, planck2013-p11, planck2013-p12, planck2013-p13, planck2013-p14, planck2013-p15, planck2013-p05b, planck2013-p17, planck2013-p09, planck2013-p09a, planck2013-p20, planck2013-p19, planck2013-pipaberration, planck2013-p05, planck2013-p05a, planck2013-pip56, planck2013-p06b, planck2013-p01a}}

\newbox\tablebox    \newdimen\tablewidth
\def\leaderfil{\leaders\hbox to 5pt{\hss.\hss}\hfil}
\def\tablenote#1 #2\par{\begingroup \parindent=0.8em
    \abovedisplayshortskip=0pt\belowdisplayshortskip=0pt
    \noindent
    $$\hss\vbox{\hsize\tablewidth \hangindent=\parindent \hangafter=1 \noindent
    \hbox to \parindent{$^#1$\hss}\strut#2\strut\par}\hss$$
    \endgroup}

\def\L2{\ifmmode L_2\else $L_2$\fi}

\def\DeltaT{\ifmmode \Delta T\else $\Delta T$\fi}
\def\deltat{\ifmmode \Delta t\else $\Delta t$\fi}
\def\fknee{\ifmmode f_{\rm knee}\else $f_{\rm knee}$\fi}
\def\Fmax{\ifmmode F_{\rm max}\else $F_{\rm max}$\fi}
\def\solar{\ifmmode{\rm M}_{\mathord\odot}\else${\rm M}_{\mathord\odot}$\fi}
\def\Msolar{\ifmmode{\rm M}_{\mathord\odot}\else${\rm M}_{\mathord\odot}$\fi}
\def\Lsolar{\ifmmode{\rm L}_{\mathord\odot}\else${\rm L}_{\mathord\odot}$\fi}
\def\inv{\ifmmode^{-1}\else$^{-1}$\fi}
\def\mo{\ifmmode^{-1}\else$^{-1}$\fi}
\def\sup#1{\ifmmode ^{\rm #1}\else $^{\rm #1}$\fi}
\def\expo#1{\ifmmode \times 10^{#1}\else $\times 10^{#1}$\fi}
\def\,{\thinspace}
\def\lsim{\mathrel{\raise .4ex\hbox{\rlap{$<$}\lower 1.2ex\hbox{$\sim$}}}}
\def\gsim{\mathrel{\raise .4ex\hbox{\rlap{$>$}\lower 1.2ex\hbox{$\sim$}}}}

\def\simprop{\mathrel{\raise .4ex\hbox{\rlap{$\propto$}\lower 1.2ex\hbox{$\sim$}}}}
\def\deg{\ifmmode^\circ\else$^\circ$\fi}
\def\pdeg{\ifmmode $\setbox0=\hbox{$^{\circ}$}\rlap{\hskip.11\wd0 .}$^{\circ}
          \else \setbox0=\hbox{$^{\circ}$}\rlap{\hskip.11\wd0 .}$^{\circ}$\fi}
\def\arcs{\ifmmode {^{\scriptstyle\prime\prime}}
          \else $^{\scriptstyle\prime\prime}$\fi}
\def\arcm{\ifmmode {^{\scriptstyle\prime}}
          \else $^{\scriptstyle\prime}$\fi}
\newdimen\sa  \newdimen\sd
\def\parcs{\sa=.07em \sd=.03em
     \ifmmode \hbox{\rlap{.}}^{\scriptstyle\prime\kern -\sd\prime}\hbox{\kern -\sa}
     \else \rlap{.}$^{\scriptstyle\prime\kern -\sd\prime}$\kern -\sa\fi}
\def\parcm{\sa=.08em \sd=.03em
     \ifmmode \hbox{\rlap{.}\kern\sa}^{\scriptstyle\prime}\hbox{\kern-\sd}
     \else \rlap{.}\kern\sa$^{\scriptstyle\prime}$\kern-\sd\fi}
\def\ra[#1 #2 #3.#4]{#1\sup{h}#2\sup{m}#3\sup{s}\llap.#4}
\def\dec[#1 #2 #3.#4]{#1\deg#2\arcm#3\arcs\llap.#4}
\def\deco[#1 #2 #3]{#1\deg#2\arcm#3\arcs}
\def\rra[#1 #2]{#1\sup{h}#2\sup{m}}

\def\dots{\relax\ifmmode \ldots\else $\ldots$\fi}
\def\WHzsr{\ifmmode $W\,Hz\mo\,sr\mo$\else W\,Hz\mo\,sr\mo\fi}
\def\mHz{\ifmmode $\,mHz$\else \,mHz\fi}
\def\GHz{\ifmmode $\,GHz$\else \,GHz\fi}
\def\mKs{\ifmmode $\,mK\,s$^{1/2}\else \,mK\,s$^{1/2}$\fi}
\def\muKs{\ifmmode \,\mu$K\,s$^{1/2}\else \,$\mu$K\,s$^{1/2}$\fi}
\def\muKRJs{\ifmmode \,\mu$K$_{\rm RJ}$\,s$^{1/2}\else \,$\mu$K$_{\rm RJ}$\,s$^{1/2}$\fi}
\def\muKHz{\ifmmode \,\mu$K\,Hz$^{-1/2}\else \,$\mu$K\,Hz$^{-1/2}$\fi}
\def\MJysr{\ifmmode \,$MJy\,sr\mo$\else \,MJy\,sr\mo\fi}
\def\MJysrmK{\ifmmode \,$MJy\,sr\mo$\,mK$_{\rm CMB}\mo\else \,MJy\,sr\mo\,mK$_{\rm CMB}\mo$\fi}
\def\microns{\ifmmode \,\mu$m$\else \,$\mu$m\fi}

\def\muK{\ifmmode \,\mu$K$\else \,$\mu$\hbox{K}\fi}
\def\microK{\ifmmode \,\mu$K$\else \,$\mu$\hbox{K}\fi}
\def\muW{\ifmmode \,\mu$W$\else \,$\mu$\hbox{W}\fi}
\def\kms{\ifmmode $\,km\,s$^{-1}\else \,km\,s$^{-1}$\fi}
\def\kmsMpc{\ifmmode $\,\kms\,Mpc\mo$\else \,\kms\,Mpc\mo\fi}
\providecommand{\sorthelp}[1]{}

\newcommand{\nh}{$N_{\textsc{H}}$}
\newcommand{\nhd}{N_{\textsc{H}}} 

\newcommand{\IRAS}{\textit{IRAS\/}}

\providecommand{\sorthelp}[1]{}

\newabbrev\ISM{Interstellar Medium (ISM)}[ISM]
\newabbrev\CSM{Circumstellar Medium (CSM)}[CSM]
\newabbrev\WNM{Warm Neutral Medium (WNM)}[WNM]
\newabbrev\WIM{Warm Ionised Medium (WIM)}[WIM]
\newabbrev\CNM{Cold Neutral Medium (CNM)}[CNM]
\newabbrev\IMF{Initial Mass Function (IMF)}[IMF]
\newabbrev\AMR{Adaptive Mesh Refinement (AMR)}[AMR]
\newabbrev\HGB{Horizontal Giant Branch (HGB)}[HGB]
\newabbrev\SFE{Star Formation Efficiency (SFE)}[SFE]
\newabbrev\TSFE{Total Star Formation Efficiency (TSFE)}[TSFE]
\newabbrev\OSFE{Observed Star Formation Efficiency (OSFE)}[OSFE]
\newabbrev\SFR{Star Formation Rate (SFR)}[SFR]
\newabbrev\YSOs{Young Stellar Objects (YSOs)}[YSOs]
\newabbrev\YSO{Young Stellar Object (YSO)}[YSO]
\newabbrev\PDF{Probability Distribution Function}[PDF]
\newabbrev\PSF{Point Spread Function}[PSF]

\makeatletter
\renewcommand\maybe@space@{%
  \maybe@ictrue 
  \expandafter   \@tfor
    \expandafter \reserved@a
    \expandafter :%
    \expandafter =%
                 \nospacelist
                 \do \t@st@ic
  \ifmaybe@ic 
    \space
  \fi
}
\makeatother

\begin{document}
\title{Interpreting the Star Formation Efficiency of Nearby Molecular Clouds with Ionising Radiation}
\author[S. Geen]
      {Sam Geen$^{1,3}$\thanks{Corresponding author: Sam Geen (sam.geen@uni-heidelberg.de)}, Juan D. Soler$^{2,3}$, Patrick Hennebelle$^{3}$ \\
{$^{1}$ Zentrum f\"ur Astronomie der Universit\"at Heidelberg, Institut f\"ur Theoretische Astrophysik, Albert-Ueberle-Str. 2, 69120 Heidelberg, Germany}\\
{$^{2}$ Max-Planck-Institute for Astronomy, K\"onigstuhl 17, 69117, Heidelberg, Germany}\\
{$^{3}$ Laboratoire AIM, Paris-Saclay, CEA/IRFU/SAp - CNRS - Universit\'e Paris Diderot, 91191, Gif-sur-Yvette Cedex, France}\\}
\date{\today}
\maketitle

\begin{abstract}

We investigate the origin of observed local star formation relations using radiative magnetohydrodynamic simulations with self-consistent star formation and ionising radiation. We compare these clouds to the density distributions of local star-forming clouds and find that the most diffuse simulated clouds match the observed clouds relatively well. We then compute both observationally-motivated and theoretically-motivated star formation efficiencies (SFEs) for these simulated clouds. By including ionising radiation, we can reproduce the observed SFEs in the clouds most similar to nearby Milky Way clouds. For denser clouds, the SFE can approach unity. These observed SFEs are typically 3 to 10 times larger than the ``total'' SFEs, i.e. the fraction of the initial cloud mass converted to stars. Converting observed to total SFEs is non-trivial. We suggest some techniques for doing so, though estimate up to a factor of ten error in the conversion.

 \end{abstract}


\section{Introduction}
\label{introduction}

\rev{In this paper we discuss how star formation relations in local Galactic clouds are set with the use of simulations of isolated star-forming regions. In particular, we measure the response of the simulated clouds to ionising radiation from massive stars, and how this alters the observed Milky Way star formation relations. We also offer some comparison between typical calculations of \SFE in numerical theory and the measurements in observations of nearby clouds.}

Stars form in molecular clouds \citep[see review by][]{Hennebelle2012}. Gas collapses into filaments and cores on the order of a freefall time. In clouds where massive star formation occurs, radiative and mechanical processes drive outflows that resist accretion around the star \citep[see review by][]{Dale2015a}. Over time, these processes disperse cloud material and suppress or halt star formation.

Quantifying star formation efficiencies in nearby molecular clouds has been the subject of much study. In one perspective, it has been proposed that star formation rates are well correlated with gas mass \citep{Lada2010} or surface density \cite{Heiderman2010} above a certain threshold, and that below this threshold the star formation rate is negligible. In a competing view proposed by \cite{Gutermuth2011}, $\Sigma_{SFR}$ is proportional to $\Sigma_{gas}^2$. \cite{Hony2015} argue that in practice distinguishing between these two arguments is difficult due to the steepness of the relations and low number statistics. The existence of a threshold for star formation has been discussed by \cite{Myers1983,Beichman1986}, who use $n_H = 10^4$ \atcc, whereas \cite{Onishi1998,Johnstone2004} use $N_H = 10^{22}$ cm$^{-2}$, equivalent to the limits found by \cite{Lada2010,Heiderman2010}.

There are various discussions as to why there should be a density criterion for star formation. \cite{Andre2010} argue that the presence of dense filaments is required for dense star-forming cores, which sets a density criterion for core formation. \cite{Burkert2013} argue that the density threshold for star formation is not a single value, but rather represents gas in which self-gravity has begun to dominate. \cite{Lee2016a} characterise the transition from the diffuse cloud to dense protocluster gas as a transition from infalling material to a virialised structure globally supported by a combination of turbulence and rotation. 

\cite{Krumholz2012b} argue that the star formation rate is in fact correlated to the local free-fall time, which is set by the volumetric density, not the column density. \cite{Clark2014} analyse simulation results and find no correlation between the column and volume density of gas, the latter of which sets the star formation rate. They do, however, find a link between the observed column density and the effective column density seen by the star. The link between observed star formation relations and the theory of star formation on small scales is still an open question.

The picture of star formation is complicated by the introduction of feedback cycles driven by energetic stellar events. In particular, a great deal of attention has been given to the role of ionising feedback. Numerical simulations by \cite{Dale2005,Gritschneder2009,Peters2010,Walch2012,Dale2012,Colin2013,Howard2016} confirm that ionising UV photons are able to drive outflows in molecular clouds that disperse the supply of dense gas during star formation events. Authors such as \cite{Whitworth1979,Matzner2002,Krumholz2006,Goldbaum2011,Kim2016} have constructed (semi-)analytic models including radiation feedback to study star formation efficiencies in clouds. One such model by \cite{Matzner2002} find that $10\pm5$ \% of the mass in clouds is converted to stars before ionising radiation disperses the cloud.

In our previous work \citep{Geen2015b,Geen2016}, we focussed on the link between simulations and analytic theory. Despite the complexity of the environments, we were successful in reproducing the quantitative behaviour of simple simulated HII regions with analytic theory \citep{SpitzerLyman1978,Dyson1980,Hosokawa2006,Raga2012,Tremblin2014a}. We were also able to explain the evaporation of dense clumps in our simulations using the models of \cite{Bertoldi1990}. By comparison, in this work we wish to link our simulations to observed clouds. We compare both the gas structure in observed clouds and the star formation efficiencies. The former is important because our initial conditions are highly idealised, and we cannot be certain how relevant our simulation results are to observed star-forming clouds otherwise.

The goal of this paper is to compare the results of observational studies of local clouds with projections of simulated star forming regions. In particular we wish to understand how ionising radiation sets star formation relations in simulations, and provide insight into how observed star formation efficiencies are obtained. We focus on the relation of \cite{Lada2010}, who measure total \YSO numbers and gas masses. By comparison, \cite{Heiderman2010} use surface densities, which we cannot reproduce accurately with our limited \YSO mass resolution. 

\rev{We stress that this comparison is for local Milky Way clouds, where fully resolved gas structures and stellar populations are available. \cite{Koepferl2016a,Koepferl2016b,Koepferl2016c} give detailed techniques for calculating star formation rates and efficiencies for cases where the cloud is not resolved and reliance on tracers is necessary, such as in extragalactic studies \citep[e.g.][]{Gao2004}.}

\rev{In this paper we are able to discuss the role of ionising radiation in producing these observed star formation relations, and compare them to typical \SFE measurements in theoretical work. The closest previous study looking at this problem, \cite{Clark2014}, omitted energetic stellar feedback processes and thus ended its analysis at a \SFE of 10\%. We purposefully use a model for the emission rates of photons that does not depend on the stochasticity of the \IMF in clusters below masses of $10^4$ \Msolar, in order to make our results simpler and easier to interpret. Similar studies of ionising radiation in clouds with varying initial properties have been performed by, e.g., \cite{Peters2010,Dale2012,Howard2016}, although using different methods and initial conditions, and without close comparison Milky Way clouds. Although the focus of this work is in using simulations to interpret observed clouds, we discuss the theoretical implications at the end of the paper.}

In Section \ref{simulations} we introduce the simulations used in this study. In Section \ref{results} we discuss our results. We begin by discussing the post-processing techniques used to approximate observations of our simulated clouds. We then present the total star formation efficiency of the clouds over their lifetime. We extend this to an estimate for the observed star formation efficiency of these clouds. We analyse the density distribution of these clouds in comparison to local clouds, and estimate the likelihood of observing each cloud at each point over its lifetime. We then discuss the implications of our study, and some of the limitations of our techniques.

\section{Numerical Simulations}
\label{simulations}

\begin{table*}
\begin{center}
\begin{tabular}{llllllll}
\textbf{Cloud Name} & \textbf{$r_{ini}$} / pc $^\mathrm{a}$ & \textbf{$\Delta$x} / pc $^\mathrm{b}$ & \textbf{$t_{ff}$} / Myr $^\mathrm{c}$ & \textbf{$n_{sink}$}/cm$^{-3}~^{\mathrm{d}}$ &  & \textbf{$\Sigma$} / \Msolarpc $^\mathrm{e}$ & \\
 & & & & & min & median & max \\
\hline
L (Most Diffuse) & 7.65 & 0.03 & 4.22 & $9.95\times10^5$ & 36.0 & 41.8 & 45.1 \\
M (Fiducial) & 3.40 & 0.026 & 1.25 & $1.25\times10^6$ & 74.5 & 75.6 & 87.0 \\
S (More Compact) & 1.9 & 0.014 & 0.527 & $3.98\times10^6$ & 160.8 & 175.0 & 185.5 \\
XS (Most Compact) & 0.85 & 0.0066 & 0.156 & $2.01\times10^7$ & 540.7 & 545.2 & 701.0 \\
\end{tabular}
\end{center}
  \caption{Cloud properties for each of the sets of initial conditions.
$^\mathrm{a}$ Initial cloud radius, excluding envelope.
$^\mathrm{b}$ Maximum spatial resolution.
$^\mathrm{c}$ The global freefall time of the cloud. Equal to $t_{relax}$, the length of time for which the cloud is relaxed (see Section \ref{simulations}).
$^\mathrm{d}$ Density threshold for sink formation and accretion, equivalent to the Jeans density in cold gas. Clumps are identified above a threshold of 0.1 $n_{sink}$.
$^\mathrm{e}$ Mean surface density for pixels above $A_=0.1$ taken at $t_{ff}$. The minimum, median and maximum values are given for values calculated along the three Cartesian axes.}
\label{simtable} 
\end{table*}

In this section we review the simulations used in this paper. We simulate a set of isolated turbulent molecular clouds with sink particles representing clusters of stars. These particles emit ionising photons that heat the gas they encounter. We use the radiative magnetohydrodynamic Eulerian \AMR code \textsc{RAMSES} \citep{Teyssier2002,Fromang2006,Rosdahl2013}.

\subsection{Initial Conditions}

We run simulations of four clouds with varying mean density, each with initial gas mass $10^4$ \Msolar. These clouds are labelled ``L'',``M'',``S'' and ``XS'', in increasing order of initial compactness. We summarise these initial conditions in Table \ref{simtable}. The last three clouds have identical initial conditions to the Fiducial, More Compact and Most Compact clouds in \cite{Geen2015b}. Cloud L is included in this work to increase the range of the study. The clouds have an initially spherically symmetric structure, with an isothermal profile out to $r_{ini}$ and a uniform sphere of radius $2~r_{ini}$ outside that with 0.1 times the density just inside $r_{ini}$. The total box length in each dimension is 48 times $r_{ini}$ in each simulation. We apply an initial velocity field with a Kolmogorov power spectrum ($P(k) \propto k^{-5/3}$) with random phases such that the cloud is in approximately virial equilibrium.

\subsection{Numerical Setup and Sink Formation}

Each simulation has a root grid of $128^3$ cells. We fully refine a sphere of diameter half the box length, encompassing the initial cloud, for two further levels. Gas that exceeds the Jeans criterion by a factor of 10 anywhere in the simulation volume is allowed to refine up to a total of 4 additional levels above the base grid (5 levels in the L cloud). We ``relax'' each cloud by halving the gravitational forces for one freefall time $t_{ff}$ in order to allow the density and velocity fields to couple.

The sink formation recipe we use is described in detail in \cite{Bleuler2014,Bleuler2014a}. We identify clumps above a threshold given by 10\% of the Jeans mass in cells at the highest refinement level. If a clump exceeds ten times this threshold, we form a sink particle. Each sink particle accretes 90\% of the mass above the sink formation density threshold at each timestep.
 
\subsection{Radiative Transfer and Cooling}
 
We track the advection of ionising photons on the AMR grid using a first-order moment method described in \citep{Rosdahl2013}. We allow the ionising photons to interact with the neutral gas, and track the ionisation states of hydrogen and helium. 

We implement the cooling function described in \cite{Geen2016}. The cooling in neutral gas is based on \cite{Audit2005}, which includes a background heating term and a fit to \cite{Sutherland1993} above $10^4$ K. Cooling of photoionised gas is treated as in \cite{Rosdahl2013}, with a piecewise fit to \cite{Ferland2003} to describe cooling on photoionised metals.

\subsection{UV Source Properties}
 
Each set of initial conditions is run twice, once with no ionising UV photons (labelled ``NRT''), and once where ionising UV photons are emitted from sink particles (labelled ``RT''). Our resolution is not sufficient to resolve individual stars. Rather, in simulations labelled ``RT'', we impose a total hydrogen-ionising photon emission rate of $S_* = (8.96\times10^{46}/s) (M_* / \mathrm{M}_{\odot})$, where $M_*$ is the total mass in sink particles. This is calculated by Monte Carlo sampling a well-sampled stellar population using a Chabrier IMF \citep{Chabrier2003}, with hydrogen-ionising emission rates for each star from \cite{Vacca1996,Sternberg2003} \citep[see Appendix A in][for further details]{Geen2016}.

We distribute this total photon emission rate across the sink particles with a weighting $F(q m_i) / \sum_{i} F(q m_i)$, where $m_i$ is the mass of a given sink particle and $F(m_i)$ is a fit to the photon emission rate for a range of stellar masses given in \cite{Vacca1996,Sternberg2003}. $q$ is a scaling factor that we set to 0.3 as a heuristic to weight emission towards more massive clumps. We adopt this model, rather than sampling stars using a Monte-Carlo scheme, to avoid stochasticity in the emission rate sampled since this would require a larger sample of simulated clouds to reach convergence.

\section{Observations}
\label{observations}

\begin{table}
\begin{center}
\begin{tabular}{llllllll}
\textbf{Cloud} & \textbf{Distance} / pc \\
\hline
\textit{Near} & (150) \\
Taurus & 140 \\
Ophiuchus & 140  \\
Lupus & 140  \\
Chamaeleon-Musca & 160  \\
Corona Australis (CrA) & 170  \\
\hline
\textit{Mid} & (300) \\
Aquila Rift & 260  \\
Perseus & 300  \\
\hline
\textit{Far} & (500) \\
IC 5146 & 400  \\
Cepheus & 440  \\
Orion & 450  \\
\end{tabular}
\end{center}
  \caption{List of Gould Belt clouds plotted in Figure \protect\ref{cumuldens}, with distances for each cloud. This list is taken from \protect\cite{PlanckCollaborationXXXV}. The distance bins used in Figure \protect\ref{cumuldens} are given in italics, with the distance assumed when calculating the \protect\PSF for the simulation results in each bin given in brackets. We use only the Serpents-South region of the Aquila Rift from Herschel.}
\label{obstable} 
\end{table}

\begin{figure*}
\centerline{\includegraphics{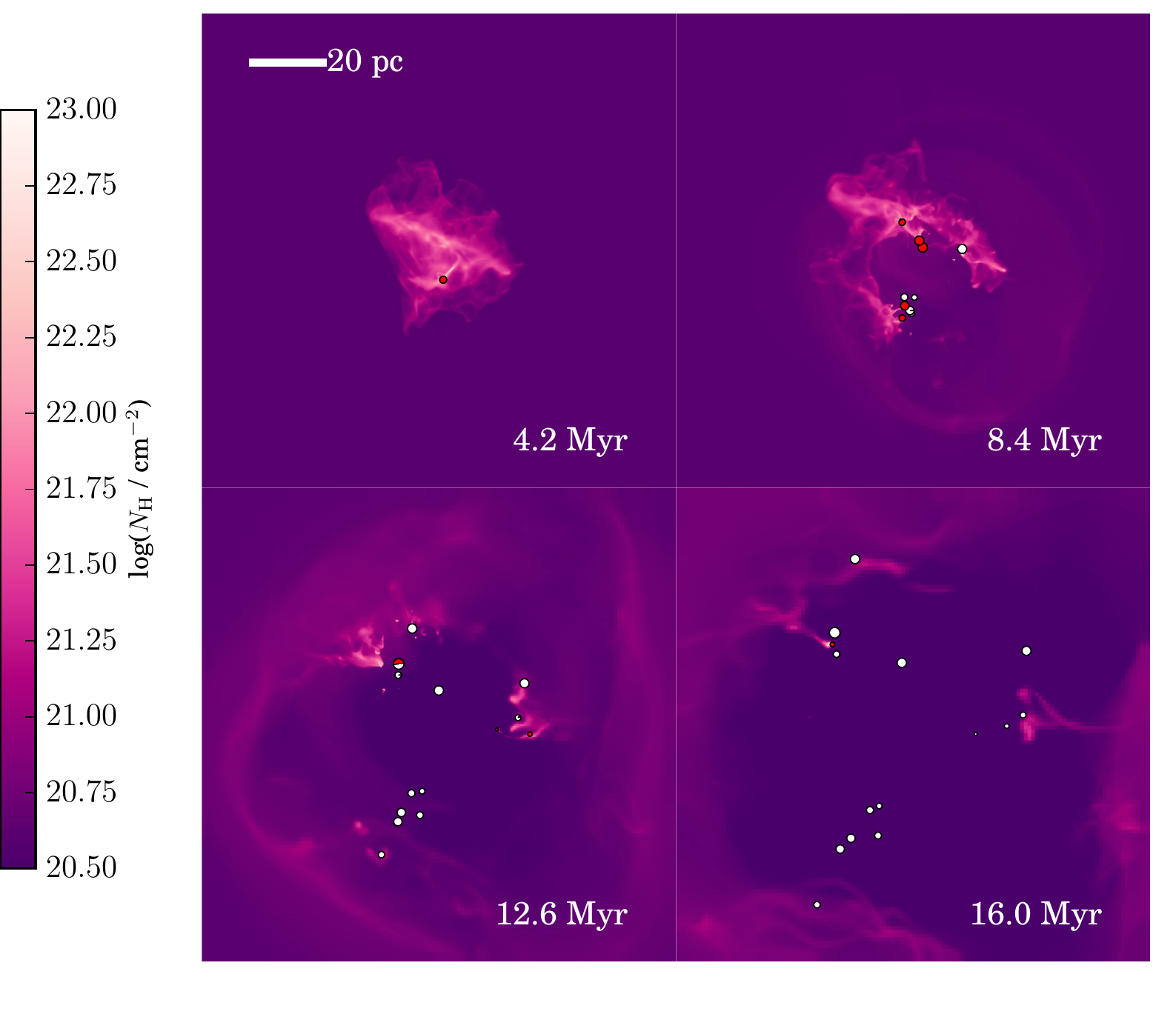}}
\caption{Sequence of hydrogen column density maps in simulation L-RT ($t_{ff}= 4.2$ Myr). Each sink particle is shown as a pie chart, whose area is proportional to the sink particle mass. The fraction shaded in red captures all of the mass accreted in the last 3 Myr, i.e. the fraction in YSOs.}
\label{images}
\end{figure*}

\rev{We use a number of observational results for which we are able to find reliable analogues in our simulation results. The most important points of comparison in this work are the \YSO count and gas masses for Gould Belt clouds given in \cite{Lada2010}. As stated in Section \ref{introduction}, this is not the only observational study to tackle this problem. However, since these authors use masses and not surface densities, and since we do not produce individual stars in our simulations, this paper provides a good point of comparison. In addition, these measures are relatively robust to systematic errors in converting the observations to physical quantities, making them ideal points of comparison with simulation results.}

\rev{In addition to the previously mentioned results, we use two sets of column density maps of the same objects, one from Planck and one from Herschel, to compare cloud structures to our simulation results on a pixel-by-pixel basis. For completeness, we summarise how each of these measurements are obtained, and how reliably they can be compared to our results.}


\subsection{YSO Counts}
\label{data:yso}

\rev{\cite{Lada2010} use a \YSO count $N_{YSO}$ in each of the Gould Belt clouds to produce \SFR estimates. To find $N_{YSO}$, they use infrared surveys from a number of sources (see the paper for a full list). Most of the clouds have been observed with the Spitzer Space Telescope, and are expected to be virtually complete\footnote{Based on tests performed using our simulations, only the densest cloud simulated in this paper has sufficient extinction to make stars in the cloud completely undetectable.}. The authors estimate that the least well sampled clouds have an error in $N_{YSO}$ of no more than a factor of 2. In these clouds they estimate the stellar ages to be $2\pm1$ Myr (see \cite{Covey2010}. They use a median stellar mass of 0.5 \Msolar to convert $N_{YSO}$ to a mass for \SFR estimates.}

\rev{In this paper, we assume a maximum \YSO age of $3\pm1$ Myr, which is approximately the lifetime of the most massive stars. Beyond this, the \YSO counts will begin to be incomplete. We use a larger \YSO age than \cite{Lada2010} since our clouds are evolved for a longer time than the given age of the observed stellar populations. The 1 Myr error is reflected in the errors in each of the figures. In addition, we shade our figures 4 Myr after $t_{ff}$ in each cloud. This is the point at which \YSO counts will start to become noticeably incomplete, and supernovae start to occur. We do not include supernovae in this paper for reasons of simplicity, and to avoid stochasticity in choices of stellar masses. Thus, any results in this region become more speculative, assuming that supernovae do indeed strongly affect the properties of the cloud.}

\subsection{Gas Masses}
\label{data:gasmass}

\rev{\cite{Lada2010} produce column density maps from extinction maps using the NICER technique \citep{Lombardi2001}. This technique measures the extinction of background stars by intervening material. Since the only parameter that affects extinction measurement is the amount of mass between the observer and the background star, it converts robustly to a column density. The authors then sum the masses of each pixel above a given extinction threshold to find a gas mass for the cloud. We define this to be $M_{A_k}$ for an extinction $A_k$. \cite{Lada2010} focus on $M_{0.1}$ and $M_{0.8}$, which are roughly analogous to all the mass in the cloud and the mass in only the densest regions, respectively.}

\rev{For each of our simulation outputs, we produce hydrogen column density ($N_H$) maps along the three cartesian axes of the simulation volume. This is just the total mass along a line of pixels, divided by the area of one pixel, converted to hydrogen number using a mass fraction $X=0.74$. We convert $N_H$ to extinction $A_k$ (in mag) using the relation $N_H = 1.67 \times 10^{22} A_k$ cm$^{-2}$ mag$^{-1}$, as given in \cite{Lombardi2008}. These maps for simulation L-RT are shown at intervals of $t_{ff}$ in the cloud in Figure \ref{images}.}

\rev{As in \cite{Lada2010}, we focus on masses contained within pixels above an extinction of $A_k=0.1 \pm 0.05$ and $0.8 \pm 0.1$. These errors are illustrative, and provide estimates for how the gas masses we produce should vary if the extinction thresholds are changed.}

\subsection{Planck Column Density Maps}
\label{data:columndensity}

\rev{We use total gas column density derived from the dust optical depth estimated from the Planck observations toward nearby (d < 500 pc) molecular clouds. Nominally, we} use maps of each of the Gould Belt objects listed in \cite{PlanckCollaborationXXXV} (see Table \ref{obstable}). For each object we use the dust optical depth at 353\,GHz ($\tau_{353}$) as a proxy for the total gas column density (\nh). 
The $\tau_{353}$ map is derived from the all-sky \Planck\ intensity observations at 353, 545, and 857$\,$GHz, and the \IRAS\ observations at 100$\,\mu$m (3000 GHz), through a modified black body spectrum fit, which also yielded maps of the dust temperature and of the dust opacity spectral index~\citep{planck2013-p06b}. These parameter maps were estimated at 5' resolution.

To scale from $\tau_{353}$ to \nh, following \cite{planck2013-p06b}, we adopted the dust opacity, 
\begin{equation}\label{eq:nhmap}
\sigma_{353}= \tau_{353}/\nhd = 1.2 \times 10^{-26}\,\mbox{cm}^{2}\,.
\end{equation}
Variations in dust opacity are present even in the diffuse ISM and the opacity decreases systematically by a factor of 2 from the denser to the diffuse ISM \citep{planck2011-7.12,Martin2011,planck2013-p06b}, but our results do not depend on this calibration.

The maps of the individual regions are projected and resampled onto a Cartesian grid with the gnomonic projection procedure described in \cite{Paradis2012a}. The present analysis is performed on these projected maps. 
The selected regions are small enough, and are located at sufficiently low Galactic latitudes that this projection does not impact significantly on our study.

\subsection{Herschel Column Density Maps}

\rev{In order to test our models of column density distributions with observations at higher angular resolutions than those possible with Planck, we use the gas column density map derived from the Herschel observations towards the Serpens South region of the Aquila Rift molecular cloud \citep[d=260 pc,][]{Prato2008a}.}

\rev{We use the 36"5 resolution NH2 column density map derived from a modified black body spectrum fit to the 70-, 160-, 250- , 350-, and 500-µm Herschel observations, described in \cite{Konyves2015a} and publicly available in the archive of the Herschel Gould Belt Survey2 \footnote{\url{http://www.herschel.fr/cea/gouldbelt}}~\citep[HGBS,][]{Andre2010} project. The instrument's resolution is below our spatial resolution at the distance of the Aquila cloud.}



\section{Results}
\label{results}

\rev{In this section we present the results of our comparisons with the observational quantities listed above. The purpose of this section is to use the simulations to describe how these observations can be physically interpreted, and how well typical simulation \SFE measurements can be compared to observed \SFE measurements}.

\rev{The broad picture of star formation in molecular clouds is well established. Stars form from dense cores in molecular clouds. The most massive of these stars inject energy into the surrounding gas in various forms, and will, depending on the properties of the cloud and stars, end star formation in the cloud and inject energy, mass and radiation in the surrounding medium.}

\rev{In our simulations, ensembles of stars are represented as sink particles. Sink particles emit radiation corresponding to the emission rate of a typical \IMF (see Section \ref{simulations}). This radiation heats the gas to $\sim10^4$ K, causing the surrounding material to expand and push away infalling matter.} This prevents further accretion onto this sink, as well as nearby sinks. Eventually the entire cloud is dispersed, halting star formation globally in the cloud. 

There are a number of separate star-forming volumes in our cloud, represented by clusters of sink particles. In panel 2 of Figure \ref{images} at 8.4 Myr, these volumes are found on the edge of the central ionisation front. This might imply that star formation is triggered by the presence of an ionisation front. However, since our star formation efficiency is reduced by the presence of star formation, we invoke the suggestion of \cite{Dale2015} that these locations would form stars anyway in the absence of an ionisation front. We note that there is no ``central'' source for the ionising radiation in panel 2. Rather, the sources of ionising radiation follow the expanding shell via gravitational attraction, and the star cluster becomes unbound by 16 Myr.

\subsection{Comparisons with Observed Cloud Structure}
\label{comparisonsclouds}

\begin{figure*}
\centerline{\includegraphics[width=0.98\hsize]{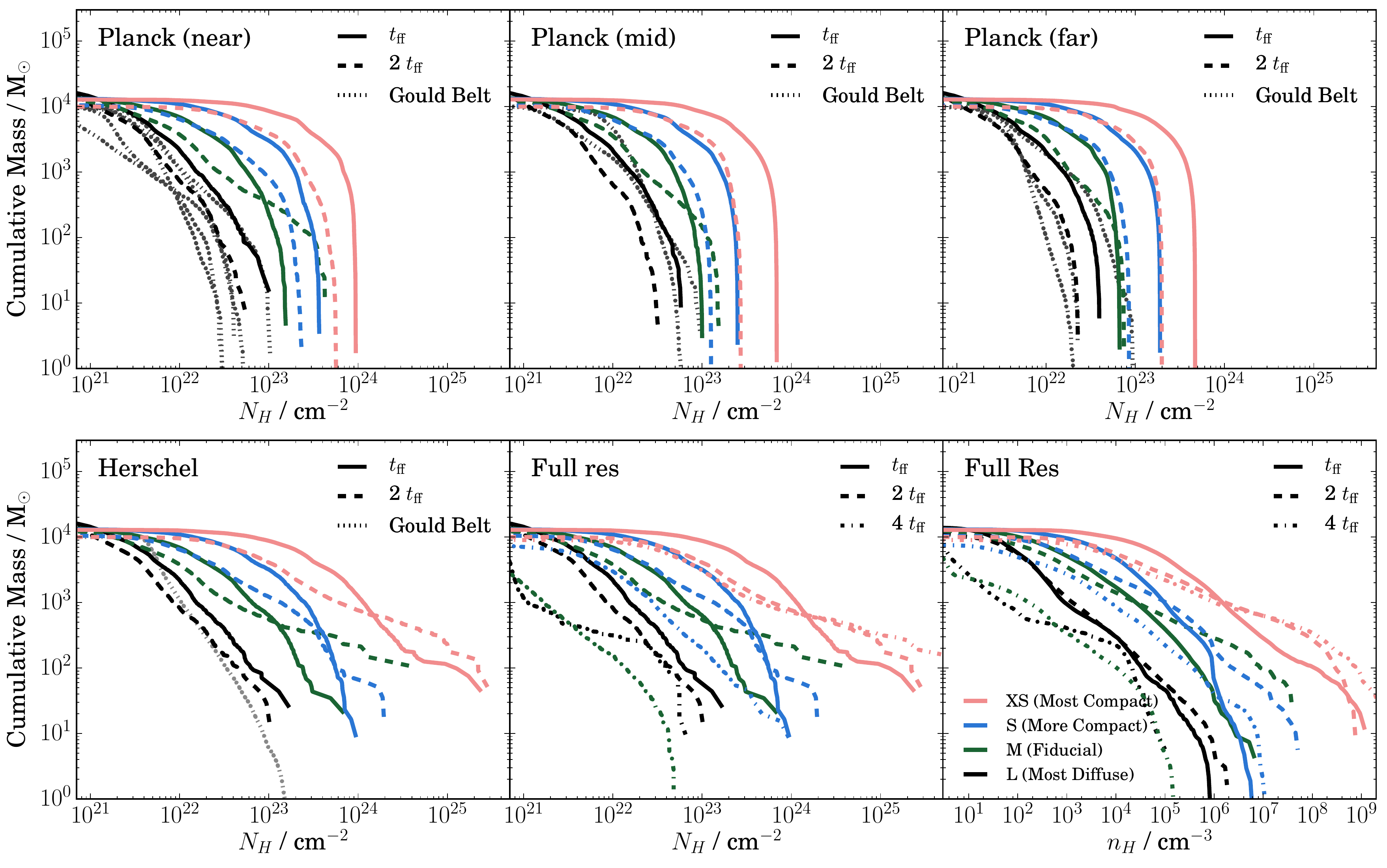}}
\caption{Cumulative density distributions in each of the simulations including ionising radiation, i.e. L-RT, M-RT, S-RT and XS-RT, with comparison to Planck and Herschel maps of Gould Belt objects (see Table \protect\ref{obstable} and Section \protect\ref{observations}). The top panels show cumulative column density distributions of our simulations at $t_{ff}$ (solid line), and 2 $t_{ff}$ (dashed line), and Planck data in groups of nearby (left, 150 pc), mid-range (centre, 300 pc) and far (right, 500 pc). Planck data is given as a grey dotted line. Each of the simulation results in the top panels is convolved with the Planck \protect\PSF, with FWHM of 5'. In each panel we assume our simulated clouds are at the distance given previously. In the left lower panel, we perform the same comparison with the Aquila region in Herschel. The FWHM here is 0.2', which gives a \protect\PSF smaller than our simulations' resolution (see Section \protect\ref{comparisonsclouds}). In the centre lower panel, we show the cumulative column density distribution of our simulations with no \protect\PSF applied. In the right lower panel we show the cumulative volume density distribution of our simulations. In these last two panels we also include our simulations results at 4 $t_{ff}$ (dash-dot line), or the final output in the simulation, whichever is earlier.}
\label{cumuldens}
\end{figure*}

In Figure \ref{cumuldens} we compare the density distribution of material in our clouds against that of the nearby Gould Belt clouds, listed in Table \ref{obstable}, using the Planck and Herschel maps. We plot the cumulative mass in bins of descending density for our simulations at $t_{ff}$ and $2 t_{ff}$ as well as the clouds described in Section \ref{observations}. We convolve our results with a \PSF corresponding to the instrument FWHM of 10' at three distance bins. These are \textit{Near} at 150 pc, \textit{Mid}-range at 300 pc and \textit{Far} at 500 pc. This removes signal from higher column densities. In the Herschel comparison, we use the Aquila cloud, which is at a distance of 260 pc. The instrument FWHM is 0.2', giving a \PSF of size 0.06 pc at this distance. This is smaller than the maximum spatial resolution in our simulations. The convolved density distributions are thus not significantly different from those taken at full resolution. The most diffuse cloud, L, matches the Planck and Herschel results most closely, although cloud M also matches some of the observed clouds at 2 $t_{ff}$.

When the results are taken at full resolution, we are able to detect denser material. Most of the clouds become denser at later times as mass accretes onto dense clumps. These densities can exceed the sink formation threshold since not all of the densest gas is accreted in a single timestep, and the freefall times at these densities are short. This pattern occurs for both column density $N_H$ and volume density $n_H$. The only cloud where this is not the case is the run L-RT. Here, HII regions driven by ionising radiation from the sinks creates cavities in the cloud at $2 t_{ff}$ (see Figure \ref{images}), reducing the number of pixels with high column densities.

Higher densities are reached in the more compact initial conditions. As well as the initial conditions being denser by construction, the maximum spatial resolution is higher due to the smaller box size. Note that clouds L and M have comparable spatial resolutions. Clouds with the density structure of clouds S and XS are not found in observed local star-forming regions (see Figure \ref{cumuldens}). In addition, these clouds already have a large quantity of gas above $A_k = 0.8$ ($\simeq N_H=10^{22}$ cm$^{-2}$) at $t=0$, requiring some mechanism to rapidly collect this much gas without forming stars. We posit that extreme environments are needed to create such objects, if indeed it is possible to collect this much dense gas without forming stars.

\subsection{Total Star Formation Efficiency}
\label{sfe}

\begin{figure}
\centerline{\includegraphics[width=0.90\hsize]{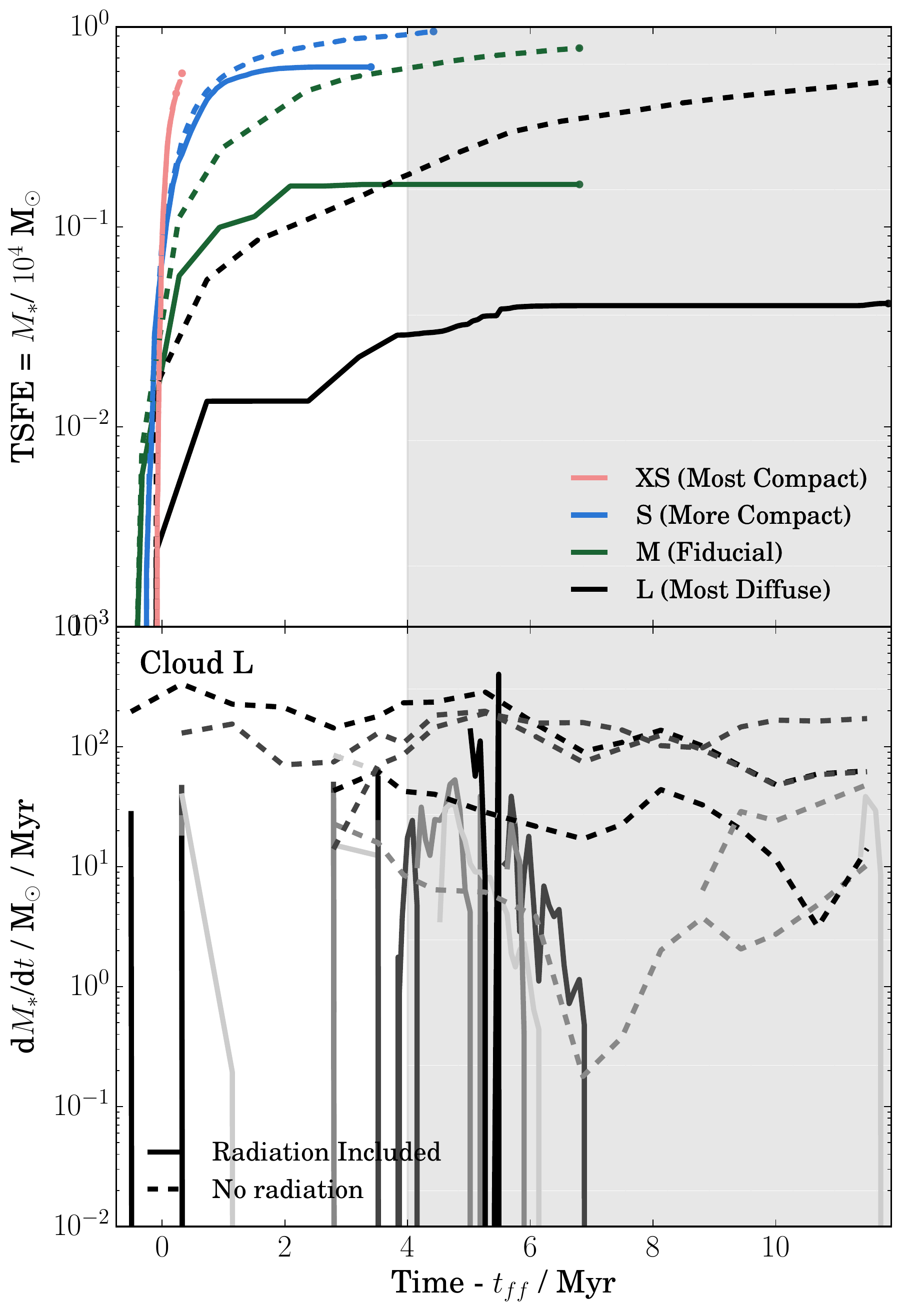}}
\caption{Top: Total star formation efficiency of each simulation over time, measured as the total mass in stars divided by the initial cloud mass. Each initial cloud density is coloured according to the legend. Simulations without radiation are shown as a dashed line, and simulations with radiation as a solid line. Bottom: The corresponding mass accretion rate over time for each particle in the L cloud setup (i.e. the most diffuse). The shades of grey alternate for each sink particle to highlight their indivdual accretion histories. \rev{$t=0$ is set to $t_{ff}$ in each cloud (see Table \protect\ref{simtable}), which is roughly the time star formation begins. The shaded region at $t=4~$Myr signifies the time at which the first supernovae can occur in a stellar population.}}
\label{totalsfe}
\end{figure}

\rev{The first quantity we use to measure \SFE is one traditionally used in in theoretical work \citep[e.g.][]{Dale2012,Colin2013,Federrath2015a,Howard2016}. This is the total fraction of the initial cloud mass $M_{ini}$ accreted onto sink particles, which we call the \TSFE. $M_{ini}$ in this work is set to $10^4$ \Msolar. We plot this quantity in Figure \ref{totalsfe}.}

In the simulations without radiation, the \TSFE tends towards 1 over several $t_{ff}$ for the cloud as a whole (see Table \ref{simtable}). With radiation, the stellar mass reaches a plateau after 1-2 $t_{ff}$. This plateau corresponds to the time at which the majority of dense gas is dispersed. In the XS simulation, this plateau is never reached due to the density of the cloud preventing ionised outflows \citep[see][]{Geen2015b}. In clouds L and M, a number of small plateaus are reached before star formation continues. This corresponds to separate star formation events within the cloud that are ended locally by ionising radiation.

In the NRT simulations, the sink particles accrete continuously over the course of the simulation until the supply of gas is exhausted. In the RT simulations, each sink particle typically accretes during a single burst anywhere between a few hundred kyr to a few Myr. After this time, radiation decouples the sink particle from the dense gas and accretion ends. The mass-weighted mean accretion time is 0.93 Myr in M-RT and 0.47 Myr in L-RT. The median accretion time is 0.63 Myr in M-RT and 0.30 Myr in L-RT. 7 of the 16 sinks in L-RT accrete for less time than the time resolution of our simulation outputs, so this figure is moderately biased towards shorter values. Feedback does not significantly shorten the accretion times in S-RT and XS-RT.

The SFE is roughly 10 times lower when ionising radiation is included than without in the L cloud, 3 times in the M cloud and closer to unity in S and XS. The XS simulations were not completed due to the short timestep in these simulations and their relative cost.

In Figure 9 of \cite{Dale2012}, clouds with mass $10^4$ \Msolar and radius above 10 pc do not form stars, while below 3 pc, ionising radiation is ineffective at altering the dynamics of the cloud. Our results are largely in agreement (see Table \ref{simtable} for initial radii $r_{ini}$). Clouds L and M both have $r_{ini}$ above 3 pc, and have reduced SFEs. Cloud L in particular is completely unbound by ionising radiation. Cloud S has an initial radius of 1.9 pc and has some reduction in response to ionising radiation, while Cloud XS ($r_{ini}=0.85$) does not respond to ionising radiation at all. Note that since we allow the cloud to relax with half gravity for one freefall time, these initial radii are underestimates of the effective cloud radius at the time stars begin to form.

%


\subsection{Reproducing Observed Star Formation Efficiency}
\label{osfesection}

\rev{In this section we attempt to reproduce observed star formation efficiencies from local clouds, where \YSO counts and gas columns are well resolved. We call this \OSFE, which is given by $M_{YSO}/M_{0.8}$ (see Section \ref{observations}).} This quantity is analagous to $\epsilon$ in \cite{Lada2010}, where $\epsilon=10\pm6\%$. The authors in this paper produce additionally star formation rates per star formation timescale and per freefall time, derived using $\epsilon$. However, since these are quantities with their own uncertainties\rev{, we focus on $\epsilon$ itself}. 

\begin{figure}
\centerline{\includegraphics[width=0.90\hsize]{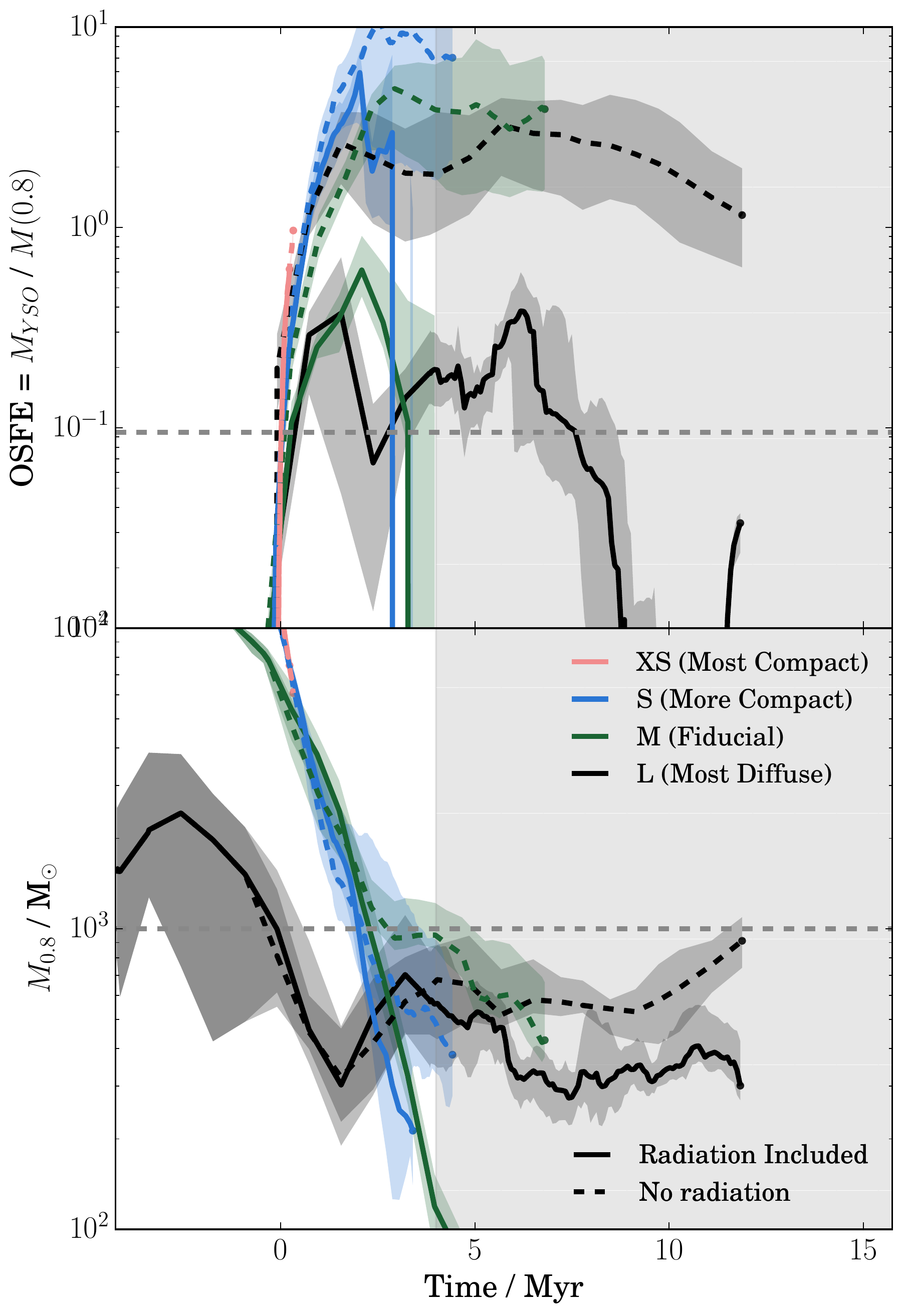}}
\caption{Top: The ``observed'' star formation efficiency \protect\OSFE in dense gas for each cloud over time, calculated as the mass in \protect\YSOs $M_{YSO}$ divided by the mass in dense gas $M(0.8)$. $M_{YSO}$ is the mass in sink particles inside contours of $A_k=0.1$ accreted in the last $t_{YSO}=3$ Myr. $M(0.8)$ is the total mass in pixels above $A_k=0.8$. Errors are given for $t_{YSO}=[2,4]$ Myr and $A_k=[0.7,0.9]$ as filled regions of the same colour. Bottom: $M(0.8)$ over time, with filled areas showing the total mass in gas between $A_k=0.7$ and 0.9. \rev{$t=0$ is set to $t_{ff}$ in each cloud (see Table \protect\ref{simtable}), which is roughly the time star formation begins. The shaded region at $t=4~$Myr signifies the time at which the first supernovae can occur in a stellar population.}}
\label{sfrvstime}
\end{figure}

In Figure \ref{sfrvstime} we plot the \OSFE as a function of dense gas mass over time for each of the simulations. \OSFE is defined as $M_{YSO}$ / $M(0.8)$. $M_{YSO}$ is the mass in sinks accreted over the last $t_{YSO}$, which we choose to be 3 Myr \citep[consistent with][noting that these authors measure the median and not the maximum \YSO age]{Covey2010,Lada2010}. As in \cite{Lada2010}, we include only sink particles inside contours of $A_k=0.1$ when calculating $M_{YSO}$. Note that \cite{Lada2010} measure the number of \YSOs, $N_{YSO}$, not the mass. They convert these using a median \YSO mass 0.5 \Msolar, i.e. $M_{YSO} = 0.5$ \Msolar $N_{YSO}$. $M(0.8)$ is the total mass in all the pixels in the gas column density maps where $A_k > 0.8$. We allow $A_k$ to vary by 0.1 and $t_{YSO}$ to vary by 1 Myr, and calculate each quantity for the line of sight along each Cartesian axis. We plot the median value as a solid or dashed line, and the maximum and minimum values as a filled area with the same colour as the line.

We overplot a line at \OSFE=10\%, corresponding to the fit to local clouds given in Figure 4 of \cite{Lada2010}. Without ionising radiation, the \OSFE lies well above this line. The denser clouds have higher star formation efficiencies, although the variation is of the same order as the estimated errors (the shaded areas in Figure \ref{sfrvstime}). With ionising radiation, the \OSFE is reduced similarly to the fraction of reduction in the \TSFE. In other words, the \OSFE is reduced by a factor of 10 in L, 3 in M and just over unity in S. There is, however, more variation due to the age of the \YSOs and the variations in dense gas mass.

All of the initial conditions for the clouds contain some gas above $A_k=0.8$, with the three densest clouds having nearly all of their mass above this threshold. Given this, the gas is rapidly converted into stars. In the denser clouds, where the freefall time is very short, ionising radiation is unable to drive outflows within a short enough time to prevent most of the cloud from being accreted onto sinks. Only the L cloud reaches a quasi-equilibrium state in which the quantity of dense gas plateaus at around $5\times10^{2}$ \Msolar.

\subsection{Likelihood of Observing a Given SFE}

\begin{figure}
\centerline{\includegraphics[width=0.90\hsize]{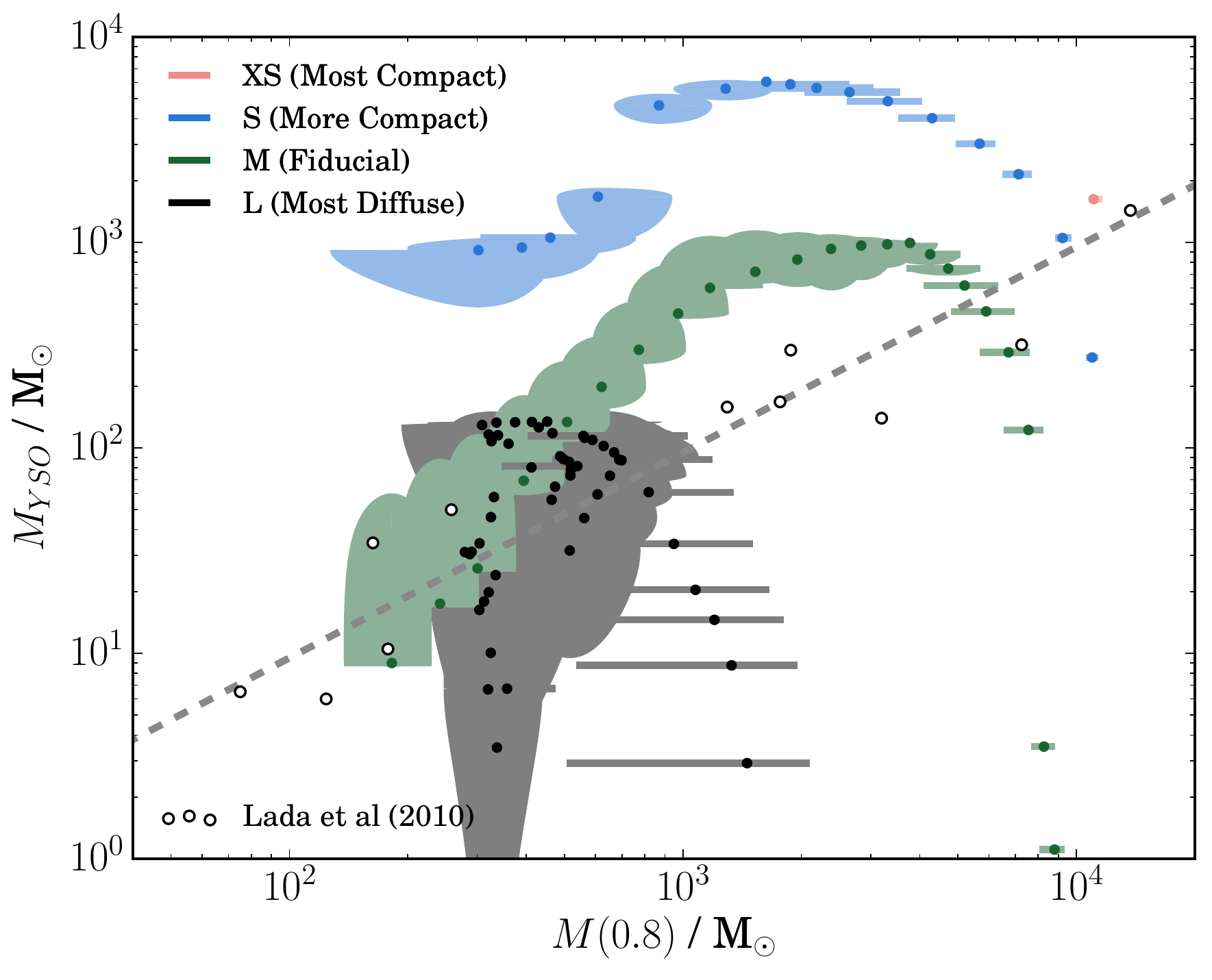}}
\caption{YSO mass, $M_{YSO}$, vs mass in gas above $A_k=0.8$, $M(0.8)$, for each simulation sampled at uniform time intervals. The YSO ages $t_{YSO}$ is 3 Myr. Errors are given for $t_{YSO}=[2,4]$ Myr and $A_k=[0.7,0.9]$. Errors in both axes are shown as circlular quadrants (in linear space), while errors present in only one axis are shown as lines in the same colour. Cloud properties are linearly interpolated between snapshots to uniform intervals of 0.2 Myr in order to illustrate the probability of observing a cloud at each given point in its evolution. The white points show data points for each cloud in Table 2 of \protect\cite{Lada2010}, while the dashed line is a fit to this data.}
\label{scattercloud}
\end{figure}

In Figure \ref{scattercloud} we plot the YSO mass $M_{YSO}$ against the mass of gas above $A_k=0.8$, $M(0.8)$, for each of the simulations with ionising radiation included. We interpolate the values of each quantity to intervals of 0.2 Myr in order to obtain the probability of observing the cloud in a given state. We overplot the same fit to \cite{Lada2010}, ${M}_{YSO}$ = 0.1 $M(0.8)$, given in Section \ref{osfesection}.

All of the clouds start at high $M(0.8)$ and low $M_{YSO}$. They rapidly reach a peak $M_{YSO}$, before tailing off to low values of $M_{YSO}$ and $M(0.8)$ as the supply of dense gas is depleted (see Figure \ref{sfrvstime}). In cloud L, in the late phase of the cloud's evolution, $M_{YSO}$ drops rapidly while $M(0.8)$ remains constant. This suggests that it is possible to accumulate supply of dense gas that is relatively quiescent in clouds that have been recently star-forming. Looking at the lower right panel of Figure \ref{cumuldens}, the maximum gas volume density at 4 $t_{ff}$ is below the threshold for forming stars in clouds L and M, despite the presence of gas above $10^{22}$ cm$^{-2}$. Since \YSOs have a maximum age ($3\pm1$ Myr) longer than the typical accretion time onto a protocluster (0.1 to 1 Myr), the presence of a \YSO is not a guarantee that stars are still actively forming in a particular volume.

Our L cloud, which most closely represents the Gould Belt clouds' density distribution (see Section \ref{comparisonsclouds}), matches the fit of \cite{Lada2010} to within the estimated spread in $M_{YSO}$ and $M(0.8)$. The initial phase of star formation is rapid, and so is less likely to be observed than the main sequence of the evolution. Finally, the late phase of cloud dispersal and minimal star formation is not sampled by the Gould Belt objects owing to their ages. The low number of nearby clouds and the limited sample of simulated clouds in this study prevents a more statistical comparison.

In addition to being less similar to local clouds, the denser clouds in our simulations live for shorter times. The most compact, XS, is only sampled for one point on our curve, by chance on the \cite{Lada2010} fit. Even if these clouds are formed in equal numbers with more diffuse clouds, we expect them to be detected in fewer numbers due to their shorter lifespans.

\section{Mapping Observed to Total Star Formation Efficiency}
\label{mappingobstototal}

\rev{In this section we discuss how well typical simluated quantities can be mapped to local observations. For the sake of clarity, we first compare stellar masses and gas masses independently.}

\subsection{YSO vs Total Stellar Mass}

\begin{figure}
\centerline{\includegraphics[width=0.90\hsize]{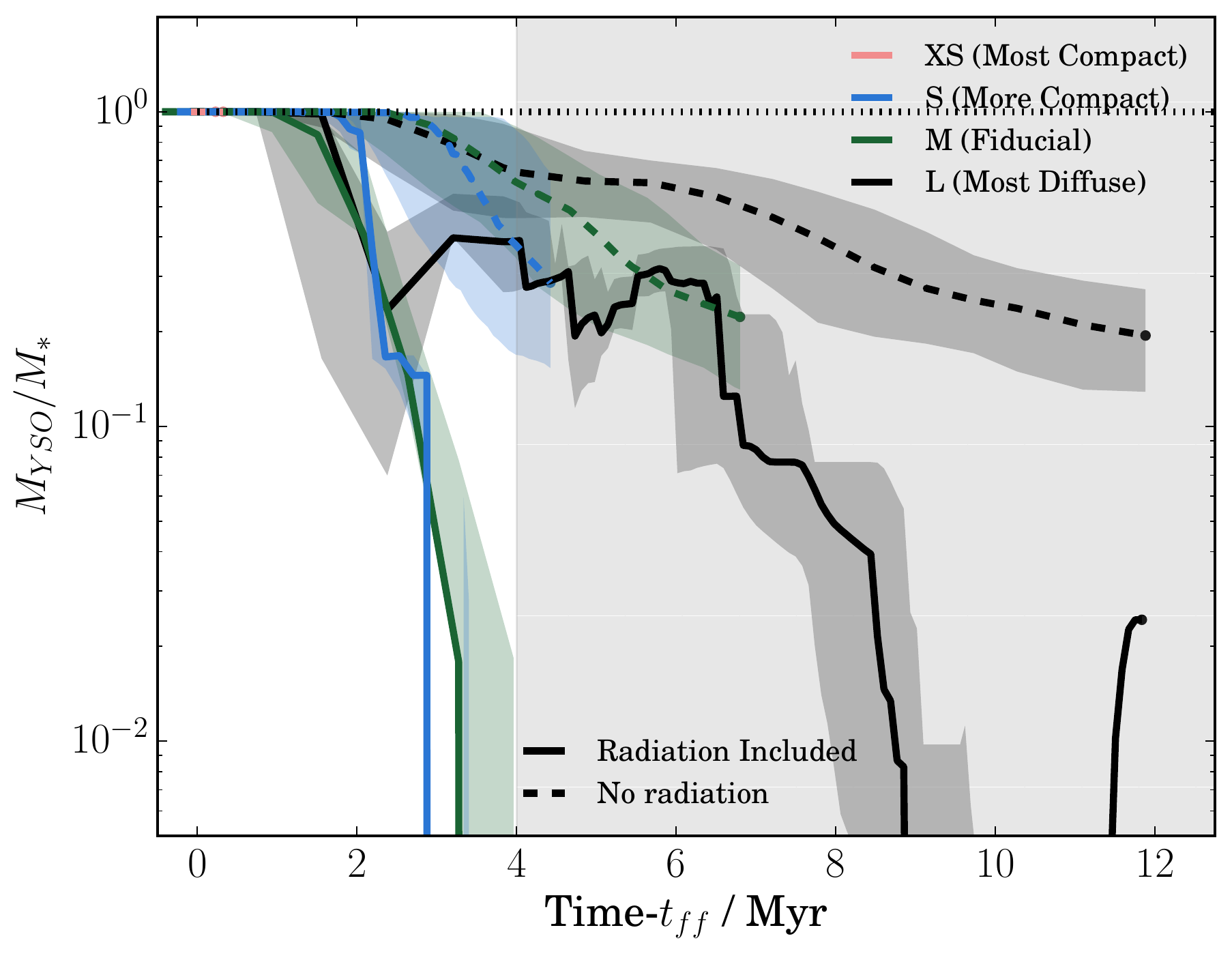}}
\caption{\rev{$M_{YSO}$ divided by total stellar mass $M_*$. $M_{YSO}$ is defined as any mass accreted by a sink particle in the last 3 Myr. $t=0$ is set to $t_{ff}$ in each cloud (see Table \protect\ref{simtable}), which is roughly the time star formation begins. The shaded region at $t=4~$Myr signifies the time at which the first supernovae can occur in a stellar population. A dotted line is drawn where $M_{YSO} = M_*$.}}
\label{obsvstotalnyso}
\end{figure}

\rev{In Figure \ref{obsvstotalnyso}, we compare the mass in recently formed stars $M_{YSO}$ with the total mass in stars formed over the lifetime of the cloud. Up to 2 Myr, the lower end of our \YSO age measurement, the ratio $M_{YSO}/M_*$ is unity. After this time, we begin to exceed the maximum age of the most massive stars in the cluster, and the ratio drops. The results of \cite{Lada2010} are most relevant in this early stage, assuming that their \YSO count is complete and no earlier phase of star formation occured. The age of stars, and loss of older stars to \YSO counts, is particularly important in the L cloud, which is most similar to the nearby Gould Belt clouds (see Section \ref{comparisonsclouds}), due to the long freefall times in the cloud.}

\subsection{Evolution of Dense and Diffuse Gas Mass}

\rev{The relation between gas mass above a certain extinction threshold $M_{A_k}$ and the total initial gas mass of the cloud is more complex. Since we use a constant $M_{ini}=10^4$ \Msolar, the behaviour of $M_{0.8}/M_{ini}$ is identical to the discussion in Section \ref{osfesection}. Initially, $M_{0.1}$ (chosen to capture all of the material above the background density) is similar to $M_{ini}$. It starts above $M_{ini}$ due to capturing part of the background in the column. Over time, the ratio drops as gas is converted to stars. It drops faster when ionising radiation is included, since this heats and disperses a proportion of the cloud mass.} 

\rev{Up to 4 Myr in the L cloud, $M_{0.1} \simeq M_{ini}$, since the conversion of gas to stars is inefficient. We thus expect \TSFE to be very similar to $M_{YSO}/M_{0.1}$ in this cloud, but only at early times. In denser clouds and later times, we expect \TSFE to map badly to \OSFE.}

\subsection{Column vs Volume Density Thresholds}

\rev{Various authors have discussed whether the mass in dense gas measured using column density thresholds on 2D maps has any relation to the mass above a volume density threshold. We refer to column thresholded masses as $M_{A_k}$ as before. Volume density thresholded masses are given as $M_{ni}$, where $i$ is given by log( $n_H$ / \atcc ), where the volume density threshold is $n_H$. This comparison has already been made in a similar study by \cite{Clark2014}, albeit without radiation. These authors find a poor match between the two quantities, with $M_{0.8}$ being much larger than $M_{n4}$.} 

\rev{In the upper panel of Figure \ref{obsvstotalvoldens}, we plot the ratio $M_{0.8}/M_{n4}$ \citep[see also][]{Clark2014}. In the lower panel we plot $M_{0.1}/M_{n1}$. The latter thresholds attempt to capture all material above the background density. After $t-t_{ff}=0$, both ratios are close to unity, until ionising radiation begins to disperse the gas. In this case, $M_{n4}$ drops faster than $M_{0.8}$ because the former is more affected by the radiation, which disperses the gas with the highest volume density, where sink particles form, first. Conversely, $M_{0.1}/M_{n1}$ drops after radiation begins to disperse the gas. This is because background material is swept into diffuse shells, which increases $M_{n1}$ while removing gas from the cloud, which reduces $M_{0.1}$.}

\rev{While there is not a single 1:1 mapping between masses derived from 2D column density maps and masses above a volume density threshold, we find that for conditions similar to local star-forming clouds, $M_{0.8}/M_{n4} = 2 \pm 1$ for the majority of the lifetime of the L cloud, and for a fraction of the lifetimes of the denser clouds. This mapping is reasonable for local clouds, which are similar in density structure to our L cloud, but as the denser clouds and the results of \cite{Clark2014} show, there is not a single mapping between the two quantities for all possible clouds.}

\subsection{Role of Extinction Threshold}

We now investigate the effect that the choice of extinction threshold $A_k$ has in setting the \SFE. In Figure \ref{akvary_obsvstotalsfe}, we plot the ratio OSFE($A_k$) / TSFE against $A_k$. There is a large degree of variation in the results at each value of $A_k$ due to variation with time. The results for XS-RT are roughly constant since nearly all of the gas is above $A_k=1$ ($N_H\simeq10^{23}$ cm$^{-2}$, see Figure \ref{cumuldens}). The median values for the other simulations follow a roughly similar trend above $A_k\simeq0.4$. This suggests that any value of $A_k$ above this limit can be used to produce a convergent star formation efficiency.

We find a systematic trend to higher values of OSFE($A_k$)/TSFE for increasing $A_k$, since as the threshold increases less gas is contained within it. Since $M(A_k)$ is typically smaller than $M_{ini}$ (see Figure \ref{sfrvstime}), the ratio OSFE($A_k$)/TSFE can be over 1 for larger values of $A_k$. 

All of the simulations display at least a factor of ten variation in OSFE/TSFE over time for a given value of $A_k$. This makes an exact conversion from OSFE to TSFE difficult. Some time variation can be accounted for by studying the time evolution of the stellar populations and HII regions \citep[see, e.g.,][]{Tremblin2014a}. Older clusters exhibit a smaller OSFE/TSFE, since the number of YSOs is truncated at 3 Myr, while the TSFE includes all stars formed by the cloud.

\begin{figure}
\centerline{\includegraphics[width=0.90\hsize]{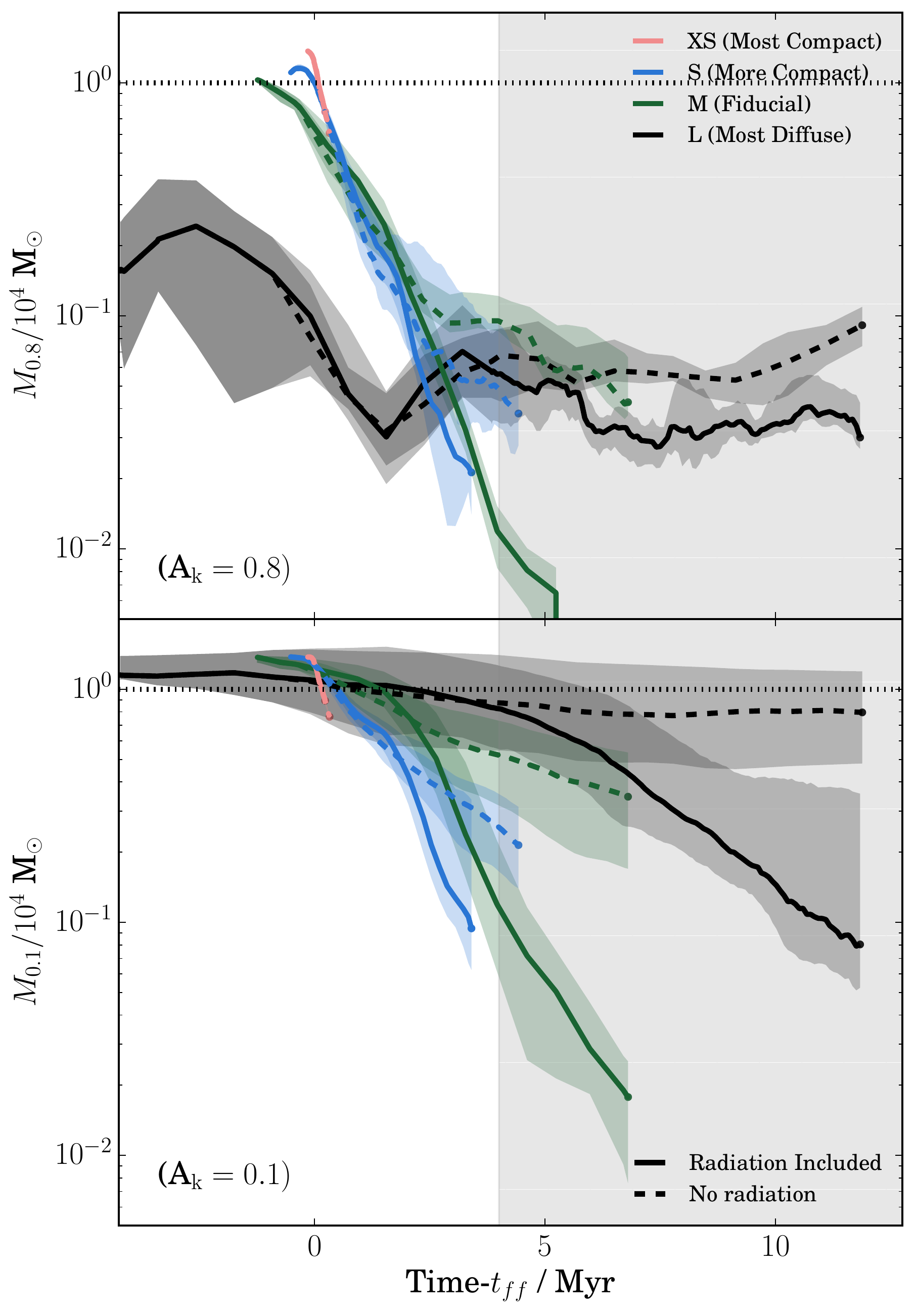}}
\caption{\rev{$M_{A_k}$ divided by total initial cloud mass $10^4~$\Msolar. $M_{A_k}$ is the mass of gas in pixels above a given extinction $A_k$. $A_k = 0.8$ in the top plot, and $0.1$ in the bottom plot. Note that some material in the envelope of the cloud and background is included in this, and so $M_{A_k}$ can be greater than $10^4~$\Msolar. $t=0$ is set to $t_{ff}$ in each cloud (see Table \protect\ref{simtable}), which is roughly the time star formation begins. The shaded region at $t=4~$Myr signifies the time at which the first supernovae can occur in a stellar population. A dotted line is drawn where $M_{YSO} = M_*$.}}
\label{obsvstotalmdense}
\end{figure}

\begin{figure}
\centerline{\includegraphics[width=0.90\hsize]{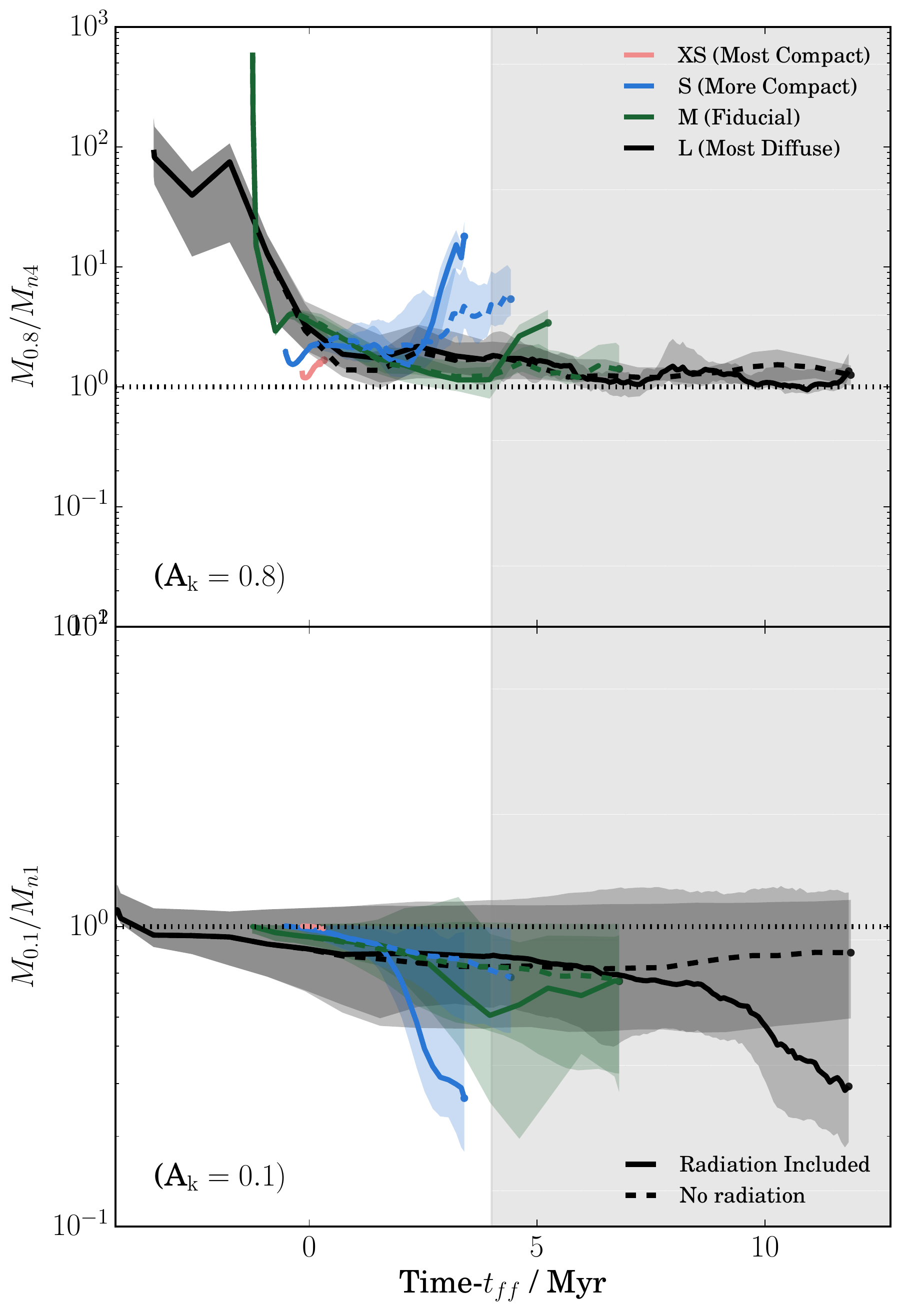}}
\caption{\rev{Figure showing the ratio between masses above a column and volume density threshold, $M_{A_k}/M_{ni}$. $M_{A_k}$ is the mass of gas in pixels above a given extinction $A_k$ (equivalent to column density $N_H$) and $M_{ni}$ is the mass above a given volume density threshold $n_H=10^i~$\atcc. The top plot shows comparisons of the dense gas only. Here we use thresholds of $A_k=0.8$ and $n_H=10^4~$\atcc, which \protect\cite{Lada2010} suggest should be equivalent. The bottom plot shows all gas identifiable as part of the cloud. Here we use $A_k=0.1$ as in \protect\cite{Lada2010}, which we compare to a threshold of $n_H=10~$\atcc (our uniform background in the simulations is 1 \atcc). $t=0$ is set to $t_{ff}$ in each cloud (see Table \protect\ref{simtable}), which is roughly the time star formation begins. The shaded region at $t=4~$Myr signifies the time at which the first supernovae can occur in a stellar population. A dotted line is drawn at unity, where both mass measurements are equal.}}
\label{obsvstotalvoldens}
\end{figure}

\begin{figure}
\centerline{\includegraphics[width=0.90\hsize]{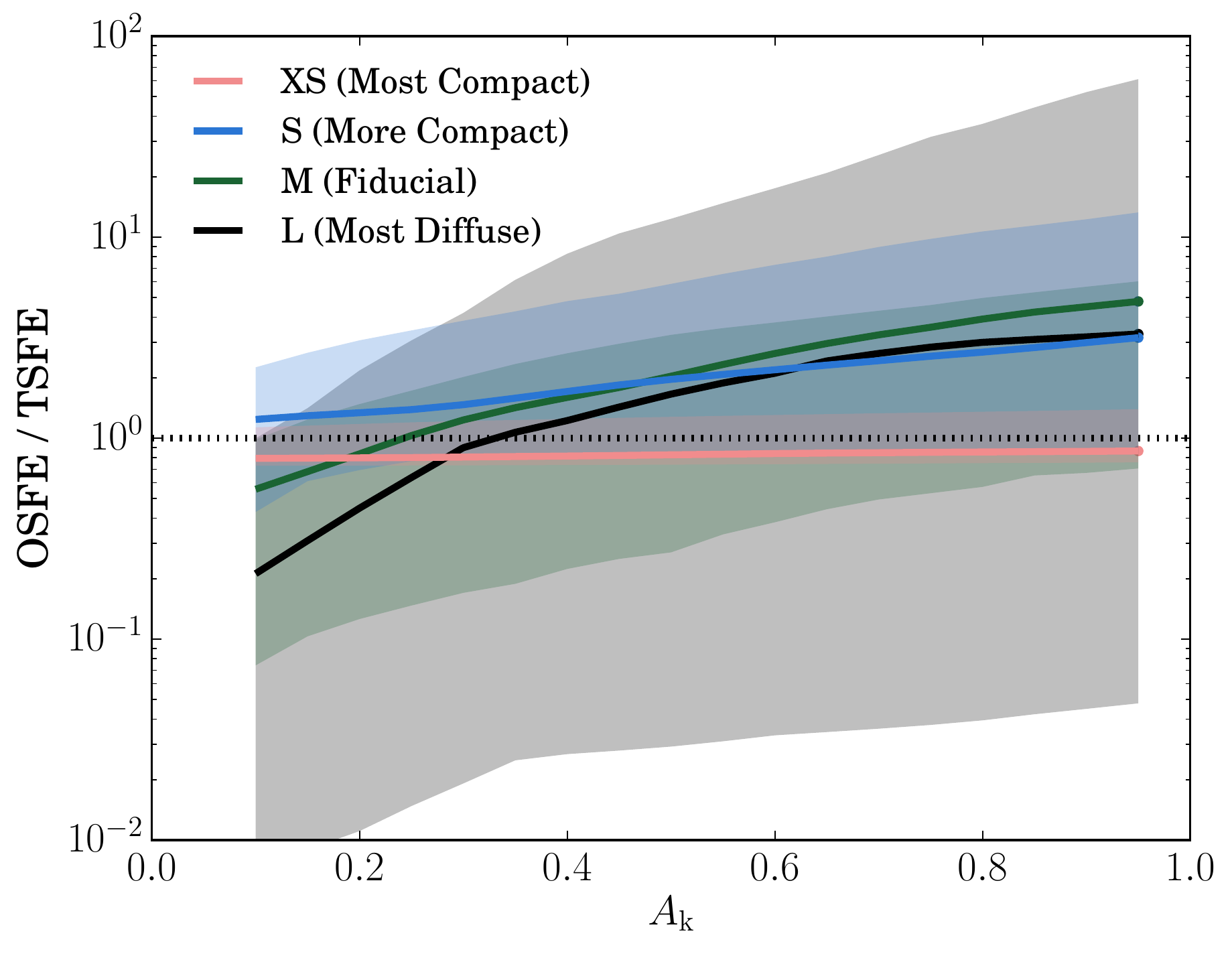}}
\caption{OSFE($A_k$) / TSFE in each of the simulations containing radiation (labelled ``-RT'') versus $A_k$. Each colour corresponds to a different simulation. The filled areas are the range between maximum and minimum values at each $A_k$ for each time, Cartesian projection and value of $t_{YSO}$ defined in Section \ref{osfesection}, excluding values where either OSFE or TSFE are zero. The solid line is the median value for each of these. The dotted line is the ratio where OSFE = TSFE.}
\label{akvary_obsvstotalsfe}
\end{figure}

\section{Discussion}

\rev{In this section we discuss the results and implications of this work. We provide points of comparison to other authors, and suggest cases where further work is needed to expand on the conclusions of this paper.}

\subsection{Relative Importance of Feedback Processes}

\rev{Producing a model for the self-regulation of star formation in clouds that accurately describes all the important processes is an extremely challenging task and one that is beyond the scope of this paper. However, it is clear that energetic stellar processes that reverse gas infall around protoclusters are an important dynamical process in setting the star formation efficiencies of molecuar clouds, regardless of the details of how such processes behave and interact with the cloud.}

\rev{A number of aspects of the problem were omitted from this project that would have increased the complexity and cost of the study. Our numerical resolution does not allow us to treat each sink as an individual star. This has two effects. Firstly, we do not resolve the full N-body behaviour of a young star cluster. Secondly, we adopt a simple prescription for the number of photons per stellar mass in the cluster. The advantage of this is that it avoids the stochasticity of small clusters with a poorly-sampled \IMF that would require an extensive parameter study of the range of possible stellar masses and photon emission rates. This undersampling of the IMF is likely to disappear for clusters above $10^4$ \Msolar \citep[see Appendix A of][]{Geen2016}.}

\rev{We also neglect processes such as stellar winds, protostellar outflows, radiation pressure and supernovae. More sophisticated models and higher resolution are needed to increase the physical fidelity of our results and allow closer comparison to observations. This is not a trivial problem, particularly if we wish to build a quantitative, descriptive model for how these processes behave. Recent 1D analytic calculations of wind and radiation pressure-dominated HII regions by \cite{Rahner2017} suggest that different processes dominate at different times and in different environments. There exists a point for which winds and radiation pressure dominate over photoionisation in HII regions. For analytic discussions of the interaction between winds and ionising radiation in HII regions, see \cite{Weaver1977,Capriotti2001}}.

\rev{Numerical simulations of \cite{Rey-Raposo2017} suggest that stellar winds do indeed reduce the star formation rates of clouds, while supernovae, which occur later, have less of an effect. \cite{Dale2014} finds less dynamical effect from winds when ionising radiation also included. The disagreement between various authors as to the importance of winds remains an outstanding problem.}

\subsection{Cloud Structure and Environment}

\rev{Molecular clouds are turbulent structures collapsing under gravity. \cite{Krumholz2007} state the importance of turbulence in lowering star formation efficiencies. In our initial conditions, we add a turbulent velocity field, but this decays over time in the absence of external sources of turbulence.}

\rev{We do not address directly in this work the origin of molecular clouds or any environmental effects that influence molecular cloud formation and destruction. Instead, we invoke starless clouds of a certain density ab initio.}

\rev{The issue of whether or not an isolated sphere is an appropriate model for a molecular cloud has been addressed by \cite{Rey-Raposo2014}, who compare clouds taken from a galactic context with isolated spheres. They find that the galactic environment is important in setting the velocity structure of the cloud. In addition, \cite{Dobbs2013} find that cloud mass evolution is strongly affected by the passage of the cloud across spiral arms.}

\rev{\cite{Ibanez-Mejia2017} use simulations of a kiloparsec-scale box and follow the collapse of clouds under the influence of supernovae and gravity. They argue that supernova driving is a secondary effect on the dynamics of the cloud, and that gravitational energy from cloud collapse, as in our isolated simulations, remains the principal driver of kinetic flows in the cloud.}

\rev{In our work, nearly all of the mass in the denser clouds (M, S and XS) is in gas identified in the dense phase (above $A_k=0.8$ or $N_H=10^{22}$ cm$^{-2}$). This means that stars are able to form rapidly before ionising radiation is able to disperse the host cloud. These artificial initial conditions cannot be linked directly to a formation mechanism, and the presence of dense gas ab initio should be treated with caution.}

\rev{Parameter studies such as this work are easier to perform in isolated spheres, since the properties of the spheres can be carefully constrained. In addition, zooming onto individual clouds presents a numerical challenge. Nonetheless, the issue of galactic environment in changing the structure and dynamics of the cloud through tidal fields, mass accretion and external feedback processes should be carefully considered.}

\subsection{Density Thresholds for Star Formation}

\rev{The existence of a density threshold for star formation is an ongoing subject of debate. In particular, the relationship between mass above a given column density and mass above a given volume density is not clear. \cite{Krumholz2012b}, for example, claim that by using a volume density threshold, a universal star formation relation can be found, and that the column density threshold is a reflection of this relation.}

\rev{In our paper, we find that during most of the lifetime of the cloud, there is a rough proportionality between gas mass above a column density threshold and gas mass above a volume density threshold for all of our clouds, with or without ionising radiation. However, \cite{Clark2014} find a larger range of conversion factors, suggesting that this is a chance occurrence. However, a wider parameter study is needed to confirm this.}

\rev{A key argument of \cite{Lada2010} is that star formation rates are linked to the mass in dense columns of gas, as opposed to the total mass of a cloud. \cite{Clark2014} suggest that there is a link between the surface density that the star ``sees'' and the projected column density to the observer. Another possibility they suggest is that denser columns are less likely to be polluted by background material. In addition to this, we find that the apparent total mass of the cloud is particularly sensitive to HII regions, which disperse material into relatively diffuse and mostly non-star-forming shells. There also does not appear to be one single density threshold that causes star formation efficiencies to converge to a single value. We find some convergence above $A_k\simeq0.4$, albeit with a large error.}

\rev{We do not explore the arguments of \cite{Heiderman2010} or \cite{Gutermuth2011} based on surface density, owing to the limited mass resolution of our cluster's sink (i.e. star) particles. \cite{Hony2015} argues that it is not possible to distinguish between models that assume a flat dependence between dense gas and stellar mass, and models that assume a power law dependence on gas density. The issue is complicated by the fact that these relations only become apparent over a number of star-forming volumes \citep{Kruijssen2014}, and break down on small scales due to incomplete sampling.}

\rev{\cite{Parmentier2017} argues that the existence of a linear relationship between dense gas mass and star formation rate requires a particular cloud structure, otherwise the relationship becomes super-linear. This is because the dense central clumps are fed slowly by infalling material, rather than collapsing quickly over one freefall time, which \cite{Krumholz2007} discount. The dynamical evolution of the cluster is also likely to confuse results in older clusters.}

\subsection{Interpreting Star Formation Efficiencies}

\rev{In this work we have measured the \SFE in a set of turbulent molecular clouds of varying initial density. We do this in two ways. The first, which we call TSFE, is typically performed on numerical simulation results. Here, we take the total fraction of the initial cloud mass converted into stars. In the second, which we call OSFE, we reproduce the measurements of \cite{Lada2010}, taken from observations of resolved, nearby clouds, which is defined as the ratio of mass in \YSOs to the mass in gas above a column of dense gas. There is not a single, clean conversion between these two measurements. Thankfully, both measurements are simple to perform on numerical simulations.}

\rev{For extragalactic clouds, where stellar populations are not resolved and the use of molecular tracers is necessary, the problem becomes more complex. \cite{Koepferl2016a,Koepferl2016b,Koepferl2016c} describe various techniques that can be used to produce accurate estimates of extragalactic star formation efficiencies. This is an important aspect, since it enables us to expand simulation results beyond local conditions and broaden comparisons with star formation across cosmic time. If indeed the structure of local clouds is similar enough that the convergent \SFE is a result of initial conditions rather than a universal star formation relation, this should be more carefully explored.} 

\rev{Our results agree broadly with those of \cite{Dale2012}, who use different numerical techniques to this work but model the same physical processes. However, we do not perform a rigorous cloud-by-cloud comparison to determine how precise this agreement is. Similarly, a link between numerical theory and simple analytic models such as those by, e.g., \cite{Matzner2002}, should be more carefully established.}

\rev{Our results also agree qualitatively with those of \cite{Colin2013}. Star formation occurs in local bursts, where individual clouds are dispersed by HII regions rather than maintaining an equilibrium of star formation. The OSFE is likely to enhance the appearance of an equilibrium state, since it uses \YSOs with a minimum lifetime of a few Myr, which smooths over much shorter bursts of active star formation. \cite{Colin2013} also argue that local variations of compactness can cause locally increased star formation rates.}

\rev{\cite{Howard2016} studies a similar problem. Curiously, they find that ionising radiation has only a limited effect on the star formation rate of the cloud. They instead argue that gravitational boundness is more important. However, their simulations do not reach the point of gas exhaustion in the cloud, suggesting that ionising radiation is less initially significant than at late times, when the cloud has been overtaken by HII regions.}

\rev{As discussed previously, the interaction between different forms of feedback in setting the \SFE is highly complicated. A convergence between numerical simulations, observational measurements and analytic theory is needed to fully grasp the problem both descriptively and quantitatively.}

\section{Conclusions}

We use radiative magnetohydrodynamic simulations with self-consistent star formation and ionising radiation from massive stars to explore the origins of observed local star formation relations. Energetic processes from massive stars drive outflows in the gas around them and end star formation in the volume around them.

We simulate a series of isolated clouds of various initial densities with and without ionising radiation. We compare the cumulative density probability distribution function (PDF) of these clouds to local clouds in the Milky Way by convolving our column density maps with the appropriate point spread function (PSF). We find that the more diffuse clouds in our sample most closely represent the observed clouds. This suggests that the densest clouds are either unphysical, or are more similar to clouds in different environments than in the Galactic neighbourhood.

We then compute the star formation efficiency (SFE) of each cloud. We define two measures of SFE. Firstly, we measure the total fraction of the initial cloud mass converted to stars, the ``Total'' SFE (TSFE). Secondly, we define an observationally-motivated SFE measurement, which we call OSFE. This is the mass in YSOs, measured as the mass recently accreted onto sink particles, divided by the gas mass in pixels above a given column density threshold. The clouds most similar to nearby molecular clouds reproduce the OSFE found by authors such as \cite{Lada2010}.

Without ionising radiation, the TSFE tends to 100\% over several freefall times. When ionising radiation is included, the TSFE drops. In the clouds most similar to local star-forming clouds, we find a TSFE of a few percent. The OSFE is typically 3 to 10 times higher than the TSFE, depending on the density threshold used. \rev{We present a number of points of comparison between observational and theoretical SFE measurements. While we observe certain trends, there is no single, clean mapping between the two methods. However, both methods are easy to compute in numerical simulation outputs, and are important points of comparison with observations of local clouds.}

\section{Acknowlegements}
\label{acknowledgements}

We thank Andreas Bleuler, Simon Glover, Sacha Hony, Ralf Klessen, Eric Pellegrini, Joakim Rosdahl and Romain Teyssier 
for their useful comments and discussions during the preparation of this work. We would also like to thank the referee for useful comments in improving the clarity of the text. This work was granted access to HPC resources of CINES under the allocation x2014047023 made by GENCI (Grand Equipement National de Calcul Intensif). This work has been funded by the the European Research Council under the European Community's Seventh Framework Programme (FP7/2007-2013). All three authors have received funding from Grant Agreement no. 306483 of this programme. SG has received additional funding from Grant Agreement no. 339177 (STARLIGHT). JS has received additional funding from Grant Agreement no. 291294.

 \bibliographystyle{mnras}
 \bibliography{Planck_bib,sfepaper}

\end{document}